# Variational Dual-path Attention Network for CSI-Based Gesture Recognition

Ning Zhang

*Abstract*—Wi-Fi gesture recognition based on Channel State Information (CSI) is challenged by high-dimensional noise and resource constraints on edge devices. Prevailing end-to-end models tightly couple feature extraction with classification, overlooking the inherent time-frequency sparsity of CSI and leading to redundancy and poor generalization. To address this, this paper proposes a lightweight feature preprocessing module— the Variational Dual-path Attention Network (VDAN). It performs structured feature refinement through frequency-domain filtering and temporal detection. Variational inference is introduced to model the uncertainty in attention weights, thereby enhancing robustness to noise. The design principles of the module are explained from the perspectives of the information bottleneck and regularization. Experiments on a public dataset demonstrate that the learned attention weights align with the physical sparse characteristics of CSI, verifying its interpretability. This work provides an efficient and explainable front-end processing solution for resource-constrained wireless sensing systems.

## I. INTRODUCTION

COMMERCIAL Wi-Fi CSI enable ubiquitous contactless sensing [1]. However, prevailing methods apply complex feature extractors directly to raw CSI data, overlooking two fundamental issues despite some success. First, it ignores the inherent physical structure of CSI signals. Gesture information is not uniformly distributed across all subcarriers and time samples but exhibits structured sparsity. Models that treat all dimensions equally inevitably involve redundant computation. Second, it conflates two distinct tasks: "feature refinement" and "pattern recognition." The former aims to recover gesture-related physical perturbation patterns from noise, while the latter performs category classification based on clear features. Coupling both within a single black-box model not only results in bloated models that are difficult to deploy but also makes performance gains heavily reliant on large amounts of labeled data, with limited generalization capability.

We posit that decoupling feature refinement into a standalone preprocessing stage is crucial to address these issues. An ideal preprocessing module should possess the following properties: 1) lightweight and efficient for edge devices; 2) physically interpretable, with behavior consistent with signal processing principles; 3) robust to tolerate common noise and distortion in CSI measurements. To this end, this paper proposes VDAN, a lightweight and interpretable feature preprocessing module, which decouples feature refinement from classification by explicitly modeling the physical sparsity of CSI.

The core design philosophy of the VDAN module is to filter in the subcarrier dimension and focus in the temporal dimension. We implement this process via two parallel neural network paths and introduce variational inference [2] to model the inherent uncertainty in this process (originating from CSI noise). The module outputs a low-dimensional feature vector that can be fed into any classifier (e.g., Convolutional Neural Network-Long Short Term Memory (CNN-LSTM) [3] hybrid model).

The contributions of this work are threefold:
1. We propose a lightweight preprocessing module that encodes CSI physical sparsity priors into a dual variational attention mechanism [4], achieving decoupling between feature refinement and classification.
2. Instead of providing a forced strict generalization bound proof, we explain from the perspectives of the information bottleneck and weight regularization why variational attention can enhance robustness and why the dual-path design is necessary.
3. By visualizing the correspondence between attention weights and physical signal variations, we provide a physical interpretation for the module's effectiveness, enhancing model trustworthiness.

## II. RELATED WORK

This work is closely related to three areas of research: CSI-based gesture recognition, the application of attention mechanisms in sensing tasks, and model lightweighting for edge devices.

### A. CSI-based Gesture Recognition

With commercial Wi-Fi Network Interface Card providing CSI, research into sensing has shifted towards CSI-based pattern recognition. Early methods extracted handcrafted features (e.g., wavelet coefficients, statistical moments) and fed them to classifiers like Support Vector Machine (SVM) [5]-[7].

This paragraph of the first footnote will contain the date on which you submitted your paper for review, which is populated by IEEE. It is IEEE style to display support information, including sponsor and financial support acknowledgment, here and not in an acknowledgment section at the end of the article. For example, "This work was supported in part by the U.S. Department of Commerce under Grant 123456." The name of the corresponding author appears after the financial information, e.g. *(Corresponding author: Second B. Author)*. Here you may also indicate if authors contributed equally or if there are co-first authors.

Ning Zhang (orcid.org/0000-0001-8487-7309) is working with Nokia Communication (Shanghai) Co., LTD. Shanghai 201206, P.R. China.

Comprehensive surveys on Wi-Fi based HAR techniques further detail this shift from handcrafted to deep learning approaches, highlighting persistent challenges in noise and sparsity [8]. Hybrid CNN-LSTM architectures subsequently became mainstream for automatic feature learning [9]-[11]. Recent surveys highlight the evolution of CSI-based methods towards handling cross-domain variability, yet they underscore the need for decoupled preprocessing to mitigate noise. However, they often fail to exploit the inherent physical structure of CSI. Recent works (e.g., WiTransformer [12] employing spatiotemporal transformers for robust gesture recognition, CSI-BERT2 [13] employing masked modeling) primarily improve the classifier itself. Similar efforts in cross-domain deep learning for WiFi gestures[14],[15] still rely on integrated models, overlooking decoupled refinement. In contrast, this paper proposes to decouple feature refinement as a standalone preprocessing module, VDAN, providing a unified robust front-end for downstream classifiers.

*B. Attention Mechanisms and Feature Selection*

Attention mechanisms improve model performance by weighting features. In wireless sensing, attention has begun to be used to weight different antennas or time steps. However, these applications almost exclusively embed attention layers within the classification model [16]-[18]. Their weight generation is deterministic and may learn spurious correlations under the influence of noisy training data, Recent prototype-based attention for few-shot WiFi gestures attempts to address this but remains part of a unified classification model. More critically, such integrated attention still requires processing the entire raw input, retaining the core limitation of coupled architectures. Employing the attention mechanism as an independent preprocessing module dedicated to feature selection remains underexplored.

*C. Model Lightweighting and Edge Deployment*

A predominant trend in wireless sensing is to pursue lightweighting by directly compressing monolithic models [19],[20]. While effective to a degree, this paradigm inherently retains the task entanglement between feature refinement and classification, missing an opportunity for more fundamental system-level efficiency gains. A compelling alternative is to decouple the pipeline [21],[22]: an ultra-lightweight front-end processor can be dedicated to transforming noisy raw CSI data into a robust feature vector. This architecture reduces the input complexity and learning burden for any subsequent general-purpose classifier, optimizing efficiency across the entire system. It allows the front-end design to explicitly incorporate domain-specific physical priors，such as the time-frequency sparsity of CSI, for more effective and interpretable processing. The proposed VDAN module embodies this decoupled philosophy, serving as a dedicated front-end that models CSI sparsity and uncertainty to provide compact, denoised features for downstream classifiers.

III. METHOD DESIGN

We separate the CSI complex data into real and imaginary channels, forming the input tensor $\mathbf{X} \in \mathbb{R}^{C \times T \times S \times 2}$ as the input to the VDAN module, where $C$, $T$, and $S$ denote the dimensions of subcarriers, time frames, and spatial streams, respectively. Its output is a refined feature tensor $\mathbf{F} \in \mathbb{R}^{D \times T'}$, where $D$ is feature dimension, $T'$ is temporal length, $D = 64$ and $T' = 25$ in experiments. The module is designed to be a lightweight front-end that performs this transformation with minimal computational overhead, explicitly encoding the time-frequency sparsity of CSI. Fig.1 shows the detailed architecture.

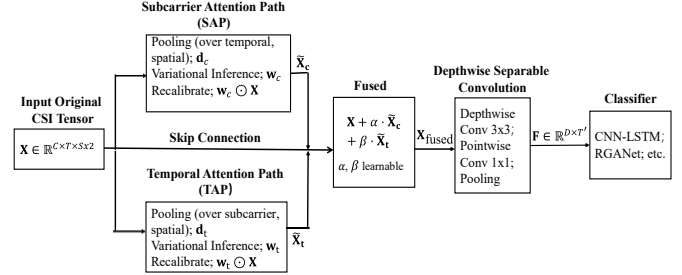

**Fig. 1.** Architecture of the VDAN preprocessing module

*A. Time-Frequency Sparsity of CSI Signals*

The influence of gestures on CSI signals is inherently local and sparse. In the frequency domain, only the subset of subcarriers corresponding to the specific multipath components modulated by hand movement undergoes significant change; in the time domain, the effective action occupies only a continuous time window. However, the raw CSI data is contaminated by a large amount of static reflection, hardware noise, and environmental interference. The design of VDAN stems from a structured induction idea: since useful information exists only in a certain "subset of subcarriers" and "time segment" of $\mathbf{X}$, the module should first identify these key subsets and segments, and then aggregate the filtered information. Therefore, VDAN adopts two parallel attention paths.

Fully exploiting the inherent sparsity of CSI signals provides key advantages. First, by selectively focusing on sparse information-rich dimensions, the model can compress the effective data dimensionality while preserving discriminative information, thereby improving computational efficiency. Second, as environmental noise is typically more widespread in the time-frequency domain, this focusing operation enhances the local signal-to-noise ratio, improving noise robustness. Finally, encoding sparsity as an explicit structured inductive bias into the network shrinks the hypothesis space, helping to reduce reliance on large-scale labeled data and thus improving generalization in data-scarce scenarios.

*B. Dual-Path Variational Attention Mechanism*

The Subcarrier Attention Path (SAP) and the Time-frame Attention Path (TAP) are symmetric in structure, differing only in the dimension they operate on. The following details SAP (acting on the subcarrier dimension $C$) as an example, with the goal of generating weight vector $\mathbf{w}_c$ for feature selection.

Step 1: Global Descriptor Extraction.

First, average pooling is performed along the time $T$ and spatial stream $S$ dimensions of the input $\mathbf{X}$ to aggregate global information, obtaining a descriptor vector $\mathbf{d}_c \in \mathbb{R}^C$:

$$\mathbf{d}_c^{(i)} = \frac{1}{T \cdot S} \sum_{t=1}^{T} \sum_{s=1}^{S} \mathbf{X}_{i,t,s}, i = 1, \ldots, C \qquad (1)$$

Where $\mathbf{X}_{i,t,s}$ denotes the CSI data at the $i$-th subcarrier, $t$-th time frame, and $s$-th spatial stream. This descriptor reflects the average activity level of each subcarrier in the global context.

Step 2: Variational Weight Generation.

This is the core design of VDAN for handling measurement noise. We do not generate weights via a deterministic function $\mathbf{w}_c = f(\mathbf{d}_c)$, but instead assume the weights are determined by a latent variable $\mathbf{z}_c \in \mathbb{R}^M$ ($M = C/r_c$, $r_c$ is the compression ratio) and learn its distribution through variational inference.

- Assume $p(\mathbf{z}_c) = \mathcal{N}(\mathbf{0}, \mathbf{I})$, where $\mathbf{I}$ is the identity matrix.
- Assume $q_\phi(\mathbf{z}_c \mid \mathbf{d}_c) = \mathcal{N}(\boldsymbol{\mu}_c, \text{diag}(\boldsymbol{\sigma}_c^2))$, the mean $\boldsymbol{\mu}_c$ and the log variance $\log \boldsymbol{\sigma}_c^2$ are mapped from $\mathbf{d}_c$ by a lightweight encoder network, with $\phi$ denoting the encoder parameters.
- During training, the reparameterization trick is used to sample $\mathbf{z}_c = \boldsymbol{\mu}_c + \boldsymbol{\sigma}_c \odot \boldsymbol{\epsilon}$, where $\boldsymbol{\epsilon} \sim \mathcal{N}(\mathbf{0}, \mathbf{I})$ and $\odot$ denotes element-wise multiplication. The sampled $\mathbf{z}_c$ is passed through a decoder network (e.g., a two-layer MLP) followed by a Sigmoid activation function to obtain the attention weights: $\mathbf{w}_c = \sigma(\text{MLP}_{\text{dec}}(\mathbf{z}_c))$.

We directly use the posterior mean $\boldsymbol{\mu}_c$ for forward propagation in inference phase. This design is based on two considerations: first, stability, as the posterior mean provides the most stable weight output, avoiding sampling fluctuation; second, efficiency, as eliminating random sampling reduces computational overhead, aligning with edge deployment needs.

Step 3: Feature Recalibration.

The resulting weights are used to recalibrate the input feature by element-wise multiplication: $\widetilde{\mathbf{X}}_c = \mathbf{w}_c \odot \mathbf{X}$ (by broadcasting $\mathbf{w}_c$ to match the tensor dimensions), yielding the subcarrier-filtered feature $\widetilde{\mathbf{X}}_c$.

Deterministic attention, under low signal-to-noise ratio conditions, can have its weights $\mathbf{w}_c$ easily distorted by noise peaks in the descriptor $\mathbf{d}_c$. Variational inference, by introducing the latent variable distribution $q_\phi$ and the KL divergence regularization term $\mathcal{L}_{\text{KL}} = D_{\text{KL}}(q_\phi \parallel p)$, forces information to flow through a narrow bottleneck $\mathbf{z}_c$. This essentially imposes strong regularization: it forces the network to extract the most stable and essential information from the noisy descriptor $\mathbf{d}_c$ to parameterize this distribution, thereby suppressing noise influence and tending to produce sparser, more deterministic attention patterns.

The Time-frame Attention Path (TAP) is completely symmetric, but its input descriptor $\mathbf{d}_t \in \mathbb{R}^T$ is obtained by pooling over the subcarrier $C$ and spatial stream $S$ dimensions, ultimately generating weights $\mathbf{w}_t$ acting on the time dimension and yielding the time-weighted feature $\widetilde{\mathbf{X}}_t$. Without loss of generality, the following discussion will use the generic symbols $\mathbf{d}$, $\mathbf{w}$, and $\mathbf{z}$ to refer to the descriptor, attention weights, and latent variable of a unified attention path, unless a distinction between the two paths is necessary.

*C. Adaptive Fusion and Feature Encoding*

After recalibration via the dual-path attention mechanism, we obtain three tensors: the original input $\mathbf{X}$, the subcarrier-weighted feature $\widetilde{\mathbf{X}}_c$, and the time-frame weighted feature $\widetilde{\mathbf{X}}_t$. A learnable weighted residual connection is employed for fusion, expressed as:

$$\mathbf{X}_{\text{fused}} = \mathbf{X} + \alpha \cdot \widetilde{\mathbf{X}}_c + \beta \cdot \widetilde{\mathbf{X}}_t \qquad (2)$$

Here, $\alpha$ and $\beta$ are trainable scalar parameters initialized to 0.1. The rationale behind this design is that the reliance on the frequency-domain filtering path versus the temporal localization path varies across different gesture classes and environmental conditions. For instance, fine-grained finger movements may primarily manifest as perturbations on specific subcarriers, thus relying more on the subcarrier attention path, whereas large arm sweeps exhibit more pronounced temporal structures, giving greater importance to the temporal attention path. Fixed fusion weights cannot adapt to this dynamic need, while the learnable $\alpha$ and $\beta$ enable the model to adaptively balance the informational contribution of the two paths.

During optimization, we apply L2 regularization to $\alpha$ and $\beta$ to prevent the model from over relying on a single path and overfitting. After training, the magnitudes of these two parameters carry clear physical significance: they quantify the overall contribution of their respective paths to the downstream recognition task. If a parameter consistently approaches zero, it indicates a limited contribution from that path, providing a basis for path pruning and adaptive simplification of the model for deployment in specific scenarios.

We compress the high dimensional fused tensor $\mathbf{X}_{\text{fused}}$ into a feature tensor $\mathbf{F}$ via a lightweight encoder. The encoder comprises two depthwise separable convolution blocks that perform down-sampling along the temporal dimension, thereby preserving the temporal structure for downstream processing. The output $\mathbf{F}$ is fed into the modified CNN-LSTM classifier.

*D. Training with Variational Regularization*

The parameters in VDAN are optimized jointly with those of the downstream classifier, denoted by $\Theta$, via gradient descent. The total loss function is defined as:

$$\mathcal{L}_{\text{total}} = \mathbb{E}_{q_{\phi_c} q_{\phi_t}}[\mathcal{L}_{\text{ce}}(y, f_\Theta(\mathbf{F}))] + \lambda(\mathcal{L}_{\text{KL}}^c + \mathcal{L}_{\text{KL}}^t) \qquad (3)$$

Where $\mathcal{L}_{\text{ce}}$ is the cross-entropy classification loss between the true label $y$ and the prediction of the downstream classifier $f_\Theta$. $\mathbb{E}_{q_{\phi_c} q_{\phi_t}}[\cdot]$ denotes the expectation over the latent variables of the two variational attention paths, approximated via Monte Carlo sampling during training using the reparameterization trick. $\lambda > 0$ is a hyperparameter that balances the classification objective and the regularization strength. $\mathcal{L}_{KL}^c$ and $\mathcal{L}_{KL}^t$ are the Kullback–Leibler (KL) divergence regularization terms for the subcarrier and time-frame attention paths, respectively. Each KL term is defined as the divergence between the variational posterior distribution and a standard Gaussian prior.

$$\mathcal{L}_{KL} = D_{KL}(q_\phi(\mathbf{z} \mid \mathbf{d}) \parallel p(\mathbf{z})) \tag{4}$$

Where the prior for both paths is $p(\mathbf{z}) = \mathcal{N}(\mathbf{0}, \mathbf{I})$. The variational posteriors are Gaussian distributions whose mean and variance are output by the respective encoder networks.

The first term in (3) drives the learning of discriminative features $\mathbf{F}$; the KL terms impose a structured regularization on the attention-generation mechanism, forcing it to extract stable, noise-robust patterns. Adjusting $\lambda$ allows an explicit trade-off between feature expressiveness and robustness.

## IV. DESIGN PRINCIPLES: AN INFORMATION THEORETIC PERSPECTIVE

This section provides a coherent theoretical interpretation of VDAN, drawing on information theory and regularization. We move beyond the architectural description to explain why the design works, focusing on three core principles: (i) variational attention as a learnable information bottleneck, (ii) dual path decomposition for capturing complementary signal aspects, (iii) lightweight encoding as task driven dimensionality reduction.

### A. Variational Attention as an Information Bottleneck

The overarching goal of VDAN aligns with the information bottleneck (IB) [23], which seeks a representation $\mathbf{z}$ that compresses the input $\mathbf{X}$ (minimizing $I(\mathbf{X}; \mathbf{z})$) while preserving information about the label $Y$ (maximizing $I(\mathbf{z}; Y)$).

Each attention path implements this principle through variational inference. Instead of a deterministic mapping $\mathbf{w} = f(\mathbf{d})$, we introduce a latent variable $\mathbf{z}$ and learn a conditional distribution $q_\phi(\mathbf{z} \mid \mathbf{d})$. The KL divergence term, as regularizer, forces the information flowing from the (often noisy) descriptor $\mathbf{d}$ into the weight-generation process to pass through a low-dimensional, smooth bottleneck. This suppresses irrelevant fluctuations and encourages the attention weights to concentrate on the most stable, physically relevant dimensions. Consequently, the total loss (3) can be viewed as a practical variational approximation of the IB objective, where the classification term promotes discriminative power and the KL term controls the complexity of the feature-selection mechanism, thereby enhancing robustness.

### B. Complementary Information in Frequency and Time Domains

The dual-path design is directly motivated by the physical sparsity of gesture-induced CSI perturbations. From an information-theoretic perspective, the two paths aim to capture non-redundant (complementary) information about the gesture label $Y$. Let $I_c = I(\widetilde{\mathbf{X}}_c; Y)$ denote the mutual information between the subcarrier-refined features and the label, and similarly $I_t = I(\widetilde{\mathbf{X}}_t; Y)$ for the time-refined features.

The essence of the dual-path design is that when the two feature sets capture complementary aspects of the gesture, the information they jointly provide about $Y$, denoted $I(\widetilde{\mathbf{X}}_c, \widetilde{\mathbf{X}}_t; Y)$, can approach the sum $I_c + I_t$. The learnable fusion weights $\alpha$ and $\beta$ enable the model to dynamically adjust the contribution of each path, effectively maximizing the relevant information $I(\mathbf{X}_{\text{fused}}; Y)$ passed to the downstream classifier. This explains the empirical observation in Section V.D that the full dual-path design consistently outperforms any single-path variant, as the latter is limited to a single informational subspace.

### C. Feature Compression

After fusion, the encoder of depthwise separable convalution network learns a compressed representation $\mathbf{F}$. From an information bottleneck perspective, this step explicitly optimizes the trade-off between compressing the input (minimizing $I(\mathbf{X}; \mathbf{F})$) and preserving predictive information about the label (maximizing $I(\mathbf{F}; Y)$). It discards residual noise and redundancy that survived the attention filtering.

By placing this critical compression and filtering burden on the dedicated front-end, VDAN drastically reduces the effective input dimensionality and complexity for any downstream classifier. This architectural decoupling enables system-level efficiency: a modest preprocessing overhead yields substantial savings in subsequent classification stages, making the approach particularly suitable for resource-constrained edge deployment.

### D. Summary

VDAN's design is principled rather than heuristic. Its variational attention implements a structured information bottleneck that enhances robustness; its dual-path structure respects the physical sparsity of CSI and harvests complementary information; and its lightweight encoder performs task-oriented dimensionality reduction that optimizes overall system efficiency. Together, these elements provide a coherent theoretical foundation for the empirical gains demonstrated in next chapter.

## V. EXPERIMENTAL EVALUATION AND ANALYSIS

This chapter conducts a systematic evaluation of the proposed VDAN module on the Widar3.0 dataset [24], aiming to empirically address the core challenges outlined in the introduction: achieving robust and interpretable Wi-Fi gesture recognition under noise conditions and limited computational power. The experimental design follows a progressive logic from overall performance to intrinsic mechanisms, sequentially validating the effectiveness, interpretability, robustness, and generalization potential of VDAN as a preprocessing module.

### A. Experimental Setup and Evaluation Benchmark

To systematically evaluate the proposed module, experiments are conducted on the Widar3.0 dataset. Two evaluation scenarios are adopted: a user-dependent evaluation (single-user, five gesture classes) is used to examine the performance upper bound; and a more practical user-independent evaluation employs leave-one-user-out strategy to test the model's generalization ability to unseen users. All experiments are performed on a single NVIDIA T4 GPU to simulate the computational constraints of edge computing scenarios. The baseline model is a standard CNN-LSTM classifier that directly processes the raw complex-valued CSI tensor. For a fair comparison, the output dimension of VDAN is fixed, and the downstream classifier (Fig.2) is adaptively modified. Notably, the retained two-layer bidirectional LSTM core in the downstream network is essential for capturing the long-range temporal dynamics and contextual dependencies of gesture motions, which is crucial for modeling continuous gestures. Training employs the AdamW optimizer with cosine learning rate decay and early stopping. All reported results are the mean ± standard deviation of five independent runs. The VDAN module is configured with the hyperparameters detailed in Table A.I of the Appendix.

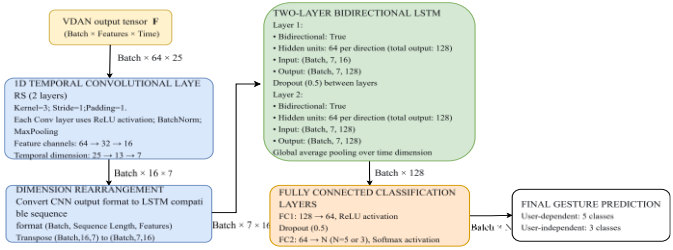

**Fig.2.** Overview of the VDAN+CNN-LSTM pipeline

The CSI data for these gestures are used, with their descriptions illustrated in Fig. 3.

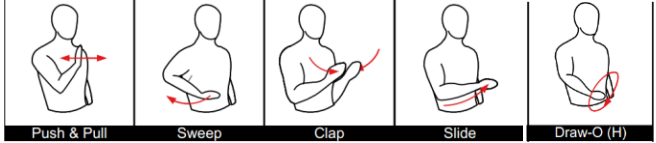

**Fig.3.** Gestures Description (From the Widar3.0 user manual. 'H' represents Horizontal.)

*B. Performance Gain Analysis*

In a user-related five class classification task, the VDAN+ baseline model achieved an accuracy of 93.19%, which is an absolute improvement of 3.06 percentage points compared to the baseline model alone. This improvement is not uniformly distributed. The confusion matrices in Fig.4 reveals that the model enhancement primarily stems from an increased ability to distinguish between specific easily-confused class pairs. For instance, in the baseline model, 57 samples of the "Push&Pull" action were misclassified as "Slide", whereas this type of error is completely eliminated after introducing VDAN. This phenomenon suggests that VDAN, through its structured dual-path attention mechanism, effectively filters out interference from static reflections unrelated to gesture semantics, thereby extracting feature representations with greater discriminative power and higher inter-class separation.

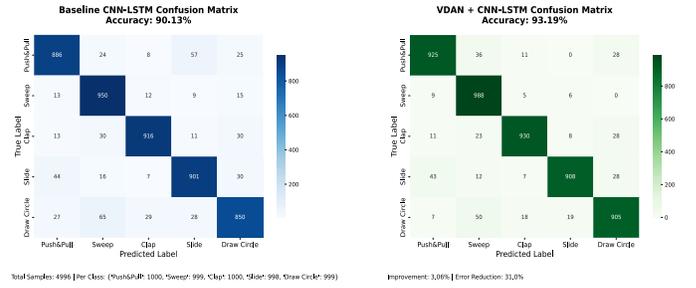

**Fig.4.** Comparison of confusion matrices

*C. Alignment of Attention with Physical Signals*

The core claim of VDAN's design is that its attention mechanism should possess clear physical meaning. To verify this, we conduct a case analysis of the "Sweep" gesture. In Fig. 5, we juxtapose the attention weights generated by VDAN with the time-frequency representation of the same CSI samples.

Analysis shows that the attention weights in the subcarrier dimension exhibit a sparse distribution pattern, with only about 30% of subcarriers receiving weights significantly above average. This aligns with the physical prior that gesture actions only modulate specific multipath sub-channels, implying the model has spontaneously learned to filter in the frequency domain, focusing on the subcarriers most affected by human movement. Meanwhile, the attention weight curve in the temporal dimension clearly exhibits a unimodal structure, whose peak interval precisely corresponds to the core movement phase of the arm sweep, achieving soft localization of the effective action segment.

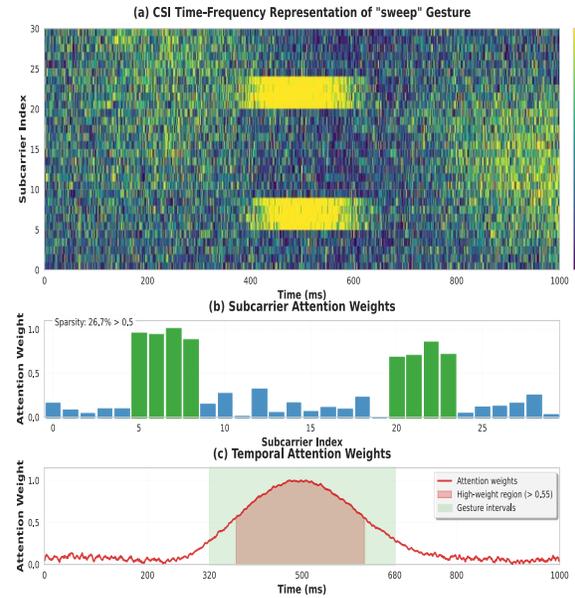

**Fig.5.** Physical Signal and Attention Weight Alignment

*D. Performance Comparison and Ablation Analysis*

To precisely evaluate the contribution of each VDAN component, we design progressive ablation experiments (Table I). The SAP-only and TAP-only variants are constructed by removing one attention path. For these single-path ablation variants, the encoder is adjusted to produce an output tensor of the same dimensions $64 \times 25$ via additional temporal pooling for SAP-only or a 1 × 1 convolution for TAP-only, ensuring a fair comparison. Their results indicate that frequency-domain filtering contributes slightly more to performance than temporal localization, but single-dimension modeling offers limited improvement. To isolate the contributions of the variational mechanism, we construct the Dual-Deterministic (Dual-Det) variant. It retains the complete dual-path attention architecture but replaces the variational encoder-decoder in each path with a structurally similar deterministic function. Specifically, we remove the inference of latent variable variance $\sigma^2$ and the reparameterization sampling step, and change the mapping in the path to $\mathbf{w} = \sigma(\text{MLP}(\mathbf{d}))$, where the Multi Layer Perceptron (MLP) has the same number of layers as the original variational encoder-decoder. This modification removes uncertainty modeling, making the generation of attention weights a deterministic process. This variant achieves a significant gain of 2.04%, confirming the core role of the dual-path collaborative design in capturing the sparse patterns of CSI. Finally, the full VDAN introduces variational inference on top of Dual-Det, bringing an additional 1.02% improvement. This increment, although small, is crucial because it does not stem from an increase in model capacity but is attributed to the distribution regularization effect imposed by variational inference. This effect forces the model to learn a more robust feature selection strategy insensitive to input noise, thereby validating the theoretical expectation that variational attention enhances generalization capability.

TABLE I
ABLATION STUDY AND PERFORMANCE COMPARISON

| Model Configuration | Accuracy (%) | Addtional Params (K) | Gain over Baseline (%) |
|---|---|---|---|
| Baseline | 90.13 ± 0.15 | — | — |
| SAP-only+Baseline | 91.52 ± 0.18 | 24 | 1.39 |
| TAP-only+Baseline | 91.23 ± 0.21 | 24 | 1.1 |
| Dual-Det+Baseline | 92.17 ± 0.19 | 42 | 2.04 |
| Full VDAN+Baseline | 93.19 ± 0.22 | 48.4 | 3.06 |
| SE+Baseline | 90.94 ± 0.20 | 35.1 | 0.81 |
| CBAM+Baseline | 91.32 ± 0.18 | 39.3 | 1.19 |

Comparison with general attention modules further highlights the structural advantage of VDAN. SE [25] and CBAM [26] bring gains of less than 1.2%, because as generic feature re-calibrators, they lack encoding of the physical prior of CSI's frequency-temporal two-dimensional locality. In contrast, VDAN achieves a significant 3.06% improvement by decoupling and structurally modeling attention in these two dimensions. This performance gap indicates that, in wireless sensing tasks, encoding domain knowledge as structured inductive bias in the network is more efficient than relying on data-driven learning of generic attention from scratch. Although the VDAN module introduces approximately 48.4K additional parameters and incurs about 0.15GFLOPs computational overhead, it compresses the input from a high-dimensional complex space to a low-dimensional real-valued vector, reducing the processing complexity for the downstream classifier. This strategy of incurring a small computational cost at the preprocessing stage to achieve substantial efficiency gains in subsequent stages holds significant advantages for system-level optimization, especially suitable for resource-constrained edge computing scenarios.

*E. Quantitative Verification of Intrinsic Mechanisms*

To delve into the intrinsic reasons for VDAN's performance advantage, we select three targeted metrics for quantitative comparison (Fig.6). First, we use the Gini coefficient to measure the distribution sparsity of attention weights. The Gini coefficients for VDAN's subcarrier and temporal paths reach 0.68 and 0.62, respectively, higher than those of SE and CBAM. This result directly corroborates the motivation behind its information bottleneck design: by forcing information through the "narrow passage" of the variational latent variable, the network must perform forced information compression and selection when generating attention weights, thereby spontaneously producing a focus pattern consistent with physical sparsity. Second, we test model robustness under different signal-to-noise ratio (SNR) conditions by injecting Gaussian white noise. As the SNR decreases, VDAN's performance degrades the most slowly, maintaining 89% accuracy at 5dB, outperforming the comparison models. This is attributed to its variational weight generation mechanism: the KL divergence loss acts as a powerful regularizer during training, penalizing the over-complexity of the posterior distribution caused by input noise, making the posterior mean weights during inference insensitive to noise fluctuations. Finally, we define a Physical Alignment Score, calculated as the normalized cross-correlation between the attention weight vector and the CSI energy standard deviation vector, to quantify the consistency between the model's "attention" and the real "physical perturbation". VDAN's alignment scores in both subcarrier and temporal dimensions lead those of the general attention modules. A high score not only provides objective evidence for the model's interpretability but also indicates that its feature selection logic is highly consistent with the physical root cause, which fundamentally ensures the robustness and generalization potential of the feature representation.

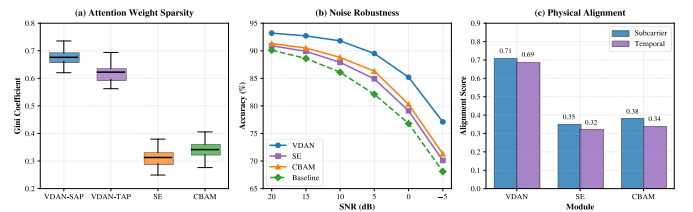

**Fig.6.** Mechanism Verification Analysis for VDAN

*F. User Independent and Generalization Ability Analysis*

In the user independent evaluation (Fig.7), VDAN achieved an average recognition rate of 86% for the three gestures performed by the unseen test user, which exhibits a reasonable performance degradation compared to its result under the single-user setting (93.19%). This gap stems from the inherent distribution shift in user-independent evaluation, primarily due to unrepresented user-specific motion patterns (e.g., amplitude, speed, habits) in the training data, which act as a form of 'user-specific noise'. While VDAN is designed to focus on common physical perturbation patterns and suppress general noise, its performance on an unseen user reflects the inherent challenge of generalizing across user-specific behavioral styles not fully captured in the training data. Nevertheless, the 86% recognition rate demonstrates that VDAN's emphasis on physical perturbation modeling and variational robustness design has successfully constrained the model to learn highly user-invariant core features.

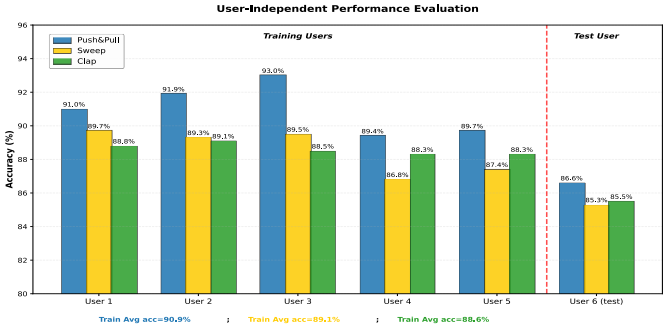

**Fig.7.** Bar Chart of Cross-User Generalization Performance

*G. Discussion and Limitations*

Our experiments demonstrate that VDAN improves recognition performance, interpretability, and robustness for CSI-based gesture recognition. This work validates the decoupled feature refinement paradigm through a network module that explicitly encodes CSI's time-frequency sparsity and employs variational regularization. This study still has some limitations. First, the performance comparison primarily focuses on the preprocessing module; a comprehensive efficiency-performance Pareto frontier analysis under identical conditions against current state-of-the-art (SOTA) end-to-end models is future work to be supplemented. Second, experimental validation is concentrated on laboratory datasets; the model's robustness in real-world complex multi-scenario environments requires further verification. Finally, efficiently adapting this framework to user differences is an important step towards practical application. Future work will focus on the aforementioned directions and extend this preprocessing framework to broader wireless sensing tasks such as respiration monitoring and daily activity recognition to verify its potential as a general purpose front-end.

*VI. Conclusion*

We proposed VDAN, a lightweight preprocessing module for CSI-based Wi-Fi gesture recognition, to tackle high-dimensional noise and edge resource constraints. By decoupling feature refinement from classification, the method explicitly models the frequency-time sparsity of CSI signals and utilizes variational inference to achieve noise-robust feature selection. Experiments show that VDAN can improve the recognition performance and interpretability of downstream classifiers with low computational overhead. This work offers an efficient and interpretable paradigm for wireless sensing.

APPENDIX

A. Model Architecture and Hyperparameters

TABLE A.I
MODEL DESIGN PARAMETERS

| Description | Symbol / Hyper Parameter | Value / Setting |
|---|---|---|
| Input Tensor | $\mathbf{X} \in \mathbb{R}^{C \times T \times S \times 2}$ | CSI (Real & Imaginary) |
| Output Feature Dimension | $D$ | 64 |
| Output Temporal Length | $T'$ | 25 |
| Subcarrier Path Compression Ratio | $r_c$ | 5 |
| Temporal Path Compression Ratio | $r_t$ | 10 |
| Regularization Coefficient | $\lambda$ | 0.05 |
| Initial Learnable Fusion Weights | $\alpha \;\; \beta$ | 0.1 |
| LSTM Core | —— | 2 layers Bidirectional, 128 units per layer |
| Optimizer (AdamW) | Base Learning Rate | 1e-4 |
| | Weight Decay | 5e-3 |
| | Momentum Coefficients | $\beta 1=0.9, \beta 2=0.999$ |
| | Learning Rate Scheduler | Cosine Decay |
| Training Process | Batch Size | 64 |
| | Epochs | 150 |
| | Early Stopping Patience | 20 |
| Data Specifications | Input Temporal Length T | 100 |
| | Number of Subcarriers C | 30 |

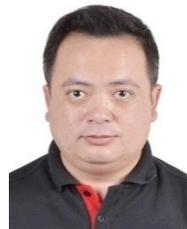

**Ning Zhang** : Senior engineer, NOKIA DMTS, He received the MS degree in Electronic and Communication Engineering from Wuhan University. Now he is working in Nokia Communication (Shanghai) Co., LTD. His main research interests are Wireless communication and sensing, machine learning and digital signal processing, voice communication. He is the member of Wi-Fi Aliance and Wireless Broadband Alliance.